\documentclass[11pt,twoside]{article}
\usepackage{asp2006}
\usepackage{psfig}
\usepackage{epsf}
\usepackage{graphics}
\usepackage{lscape}
\def\edcomment#1{\iffalse\marginpar{\raggedright\sl#1\/}\else\relax\fi}
\reversemarginpar

\begin{document}
\title{A dichotomy in radio jet orientations in elliptical galaxies}
\author{I.W.A.~ Browne and R.A.~Battye}
\affil{Jodrell Bank Centre for Astrophysics, School of Physics and Astronomy, The University of Manchester, Oxford Road, Manchester, M13 9PL}

\begin{abstract}
We have investigated the correlations between optical and radio
isophotal position angles for 14302 SDSS galaxies with $r$ magnitudes
brighter than 18. All the galaxies are identified with extended FIRST
radio sources. For passive early-type galaxies, which we distinguish
from the others by using the colour, concentration and their principal
components, we find a strong statistical alignment of the radio axes
with the optical minor axes.  Since the radio emission is driven by
accretion on to a nuclear black hole we argue that the observed
correlation gives new insight into the structure of elliptical
galaxies, for example, whether or not the nuclear kinematics are
decoupled from the rest of the galaxy. Our results imply that a
significant fraction of the galaxies are oblate spheroids, perhaps
rotationally supported, with their radio emission aligned with the
stellar minor axis.  Remarkably, the strength of the correlation of
the radio major axis with the optical minor axis depends on radio
loudness. Those objects with a low ratio of FIRST radio flux density
to total stellar light show a strong minor axis correlation while the
stronger radio sources do not. This split may reflect different
formation histories and we suggest this may be a new manifestation of
the better known dichotomy between slow rotating and fast rotating
ellipticals.
\end{abstract}

\vspace{-0.5cm}
\section{Introduction}
Searching for statistical alignments between radio and optical axes is
motivated by the desire to find a connection between radio emission
mechanism and the geometry of the host galaxy. Though there has been a
long history of searching for such alignments, generally the results
have been inconclusive (for example, \cite{Mackay1971};
\cite{Palimaka1979}; \cite{Valtonen1983}; \cite{Birkinshaw1985};
\cite{Sansom1987}). In most investigations there appears to be a slight
preponderance of objects where the radio elongation is more aligned
with the optical minor axis than the major axis. The clearest result
was obtained by \cite{Condon1991} who found that extended radio jets
in 125 UGC galaxies were preferentially aligned with the optical minor
axes of their hosts, with the effect being strongest for elliptical
galaxies. 
Recent work indicates that elliptical galaxies have complex kinematics. These
include decoupled cores in a
significant fraction of galaxies (for example,
\cite{Halliday2001}; \cite{Loubser2008}; \cite{Krajnovic2008}), and
the separation of such galaxies into two distinct classes, cored on
non-cored (\cite{Kormendy2009}) and/or fast and slow rotators
(\cite{Emsellem2007}). Therefore studying the alignment between the stellar
population and the radio emission assumes added significance.

All the historic results have been based on relatively small numbers
of objects ($\sim$ 100) but with the advent of deep radio and optical
surveys like FIRST (\cite{Becker1995}) and the SDSS (\cite{York2000}),
respectively, it is possible to construct samples with orders of
magnitude larger numbers. But going deeper in radio flux density means
that the galaxies identified with the radio sources are no longer
dominated by the ellipticals, which were the targets of early studies,
but are now a mixture of ellipticals and disk-dominated star-forming
galaxies. Since the source of the emission is fundamentally different
in the two types of galaxy one might well expect their alignment
properties also to be different. Hence we need to distinguish reliably
between early- and late-type galaxies and we do this by using SDSS
photometry. In this paper we will focus almost exclusively on radio
sources hosted by elliptical galaxies.

\section{Sample selection}

We use optical identifications of FIRST radio sources listed in the
SDSS DR6 database (\cite{Adelman2008}) which were selected using the
procedure defined in \cite{Ivezic2002}.  There are 239993 such
identifications. We have also extracted the photometric
magnitudes from SDSS ($ugriz$), the integrated flux density measured
by FIRST ($S_{\rm int}$), the position angle (PA, $\alpha$) computed from the isophotal
distribution in the $r$-band by SDSS along with the equivalent from
FIRST, the major ($a$) and minor ($b$) axes measured by SDSS and
FIRST, and $R_{50}$ and $R_{90}$ measured from the Petrosian intensity
profile model fitted to the SDSS images (\cite{Petrosian1976}). We
also use the concentration $c=R_{90}/R_{50}$ to help distinguish
elliptical galaxies from the others; for small $c$ the light
distribution has an exponential profile and fits the profile of a disk
galaxy, whereas for large values of $c$ the profile is approximated by
a de Vaucouleurs profile which fits the light distribution of an
elliptical galaxy.

For our investigation we require galaxies and radio sources for which
there are reliable measures of the PAs of extended emission. Since it
is difficult to compute the PA accurately for very faint galaxies and
very round ones, we have excluded all galaxies with $r>18$, $b/a>0.8$
in either SDSS or FIRST, and those with $a<2^{\prime\prime}$ in FIRST
(since FIRST has a $5^{\prime\prime}$ beam). This leaves a total of
14302 galaxies. We further split the sample based on photometric
determined parameters. It has been shown by \cite{Strateva2001} that
the colour defined by $u-r$ can discriminate between different
distributions in the bimodal $g-r$ vs. $u-g$ colour-colour diagram for
an earlier SDSS data release. The colour and concentration are known
to be correlated (see for example, \cite{Strateva2001}). There is a
bimodality in the density of points in the colour-colour diagram for
our sample, suggesting that there are indeed two populations, with the
redder, more highly concentrated objects being the elliptical
galaxies. In order to investigate this quantitatively we have
performed a principal component analysis (PCA) on the sample,
initially with two variables $u-r$ and $c$. The two components which
are generated by this procedure are
\begin{eqnarray}
C_1&=&0.965c-0.262(u-r)\,,\cr
C_2&=&0.262c+0.965(u-r)\,.
\end{eqnarray}
Dividing at $C_2=3.5$ gives a near optimum
separation of the two types of galaxy.

\section{Results}

\begin{figure}
\centerline{\psfig{figure=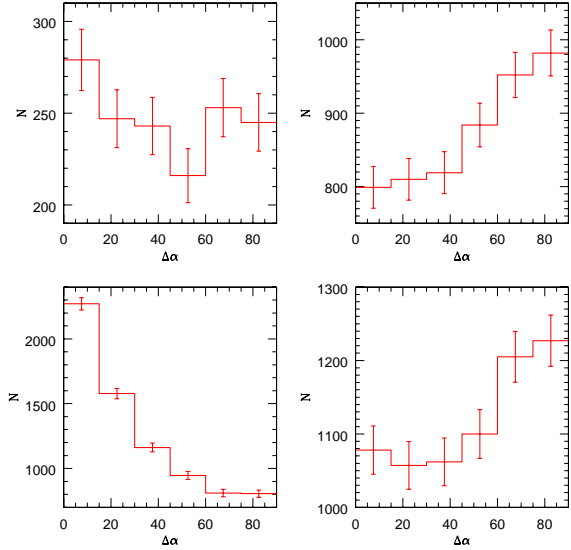,width=8cm,height=8cm}}
\caption{Splits of the data on the basis of PCA and axial ratio. The
distributions are: $C_2<3.5$ (bottom left), $C_2>3.5$ (bottom right),
$C_2>3.5$ and $b/a<0.6$ (top left) and $C_2>3.5$ and $b/a>0.6$ (top
right)}
\label{fig:hist_corr}
\end{figure}

\begin{figure}
\centerline{\psfig{figure=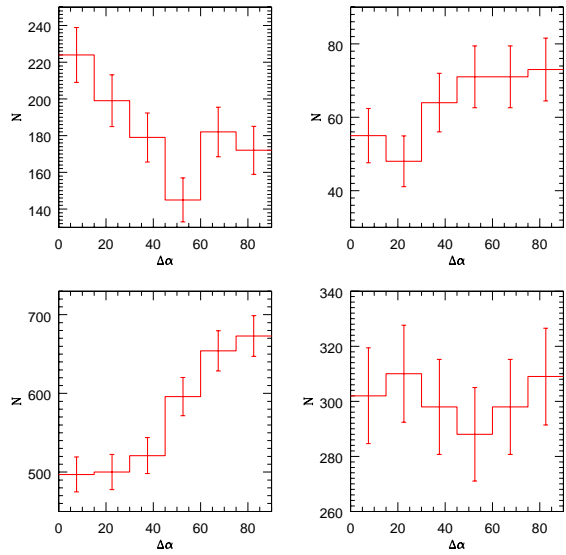,width=8cm,height=8cm}}
\caption{Histograms for  the case $C_2>3.5$ with  various splits based
on $b/a$ and $t-r$. The two plots at the bottom have $b/a>0.6$ and the
two the  top have $b/a<0.6$. The two  on the left are  radio-quiet with
$t-r>-2.5$ and the two on the right are radio-loud with $t-r<-2.5$.}
\label{fig:histsplit2}
\end{figure}

In Fig.~\ref{fig:hist_corr} we show the distribution of $\Delta\alpha$
split by principal component $C_2$ and by axial ratio $b/a$. As
expected the two types of galaxy behave differently. It is very
evident that the blue, less concentrated galaxies ($C_2<3.5$), which
broadly represent the disk-dominated, spiral population, have the
optical and radio major axes correlated, whereas the red, highly
concentrated galaxies ($C_2>3.5$), which are part of the elliptical
population, have a correlation between the optical major and radio
minor axes. The most statistically significant minor axis alignment is
obtained for a sub-sample defined by $C_2>3.5$ and $b/a>0.6$. We draw
attention to the fact that we are dealing with relatively round
galaxies.  Though the major axis alignment for the spiral galaxies is
interesting in its own right, for the rest of this contribution we
will concentrate our discussion on the early-type galaxies. An
extensive discussion of the behaviour of both types of galaxy can be
found in \cite{Battye2009}.

The bias towards $\Delta\alpha=90^{\circ}$ in the red objects with
high concentrations is highly statistically significant. Since in
these predominantly elliptical galaxies the radio-emission is almost
certainly powered by mass accretion on to a central disk/black hole
system, the radio elongation we measure will be that of the overall
spin axis of that system. Our results confirm those reported by
\cite{Condon1991} who saw a strong trend for the jet axes in UGC
elliptical galaxies to be aligned with the optical minor axes.
Simulations done by \cite{Sansom1987} for different galaxy geometries
show that strong minor axis alignments are observable only for
galaxies which are oblate spheroids.  Therefore our results imply that
there is a bias for the spin axis of the central engine to be aligned
with the minor axes of galaxies and also that these galaxies are
predominantly oblate spheroids, something supported by the analysis of
galaxy shapes (\cite{Padilla2008}). The radio minor axis alignment is
something one might expect to see if the galaxies were rotationally
supported and the black hole accretion disk axes were aligned with the
overall galaxy rotation axes. One should contrast what we see here in
elliptical galaxies with that seen in Seyfert galaxies where there is
only a weak relationship between jet axis and the axis of the host galaxy
disk (\cite{Kinney2000}; \cite{Gallimore2006}; \cite{Raban2009}).

We have divided the sample further by the ratio of radio 
to optical flux density expressed as $t-r$ where r is the r magnitude
and t is given by:
\begin{equation}
t=-2.5\log_{10}\left(S_{\rm int}\over 3631{\rm Jy}\right)\,.
\end{equation}
What is most remarkable is that the strong $\Delta\alpha=90^{\circ}$ bias in
the elliptical galaxies is
confined to the radio quieter subset of these objects as is clearly evident from Fig.\ref{fig:histsplit2} 
The division at $t-r=-2.5$ produces sub-samples with
a ratio in numbers of around 2:1 and was not chosen {\it a priori} to
have any particular physical significance for the elliptical galaxy
population. However, the division seems to have 
physical significance, indicating that there are two distinct types of
object within the elliptical population exhibiting quite different
behaviour. We speculate that the galaxy shape in the quieter objects
is fixed by rotation and that jets emerge along the stellar rotation
axis. We emphasize that the two radio populations we talk about here
have nothing to do with the traditional FR1/FR2 radio morphological
division that for an L$^{*}$ galaxy occurs at around $t-r=-7$ not
$t-r=-2.5$. Virtually all our galaxies are in the FR1 range.

\section{Discussion; two varieties of elliptical galaxies}

\begin{figure}
\centerline{\psfig{figure=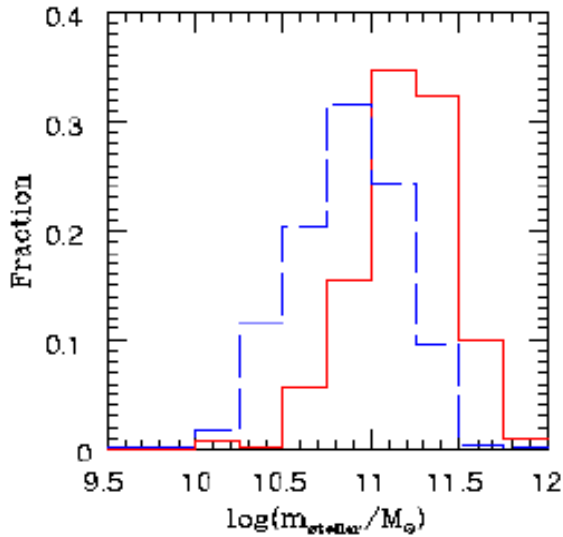,width=8cm,height=8cm}}
\caption{Plot of the stellar masses of the sample split by the ratio of radio to optical luminosity. The lower luminosity objects are indicated by the dashed line and the higher by the continuous line.}
\label{fig:mass}
\end{figure}

We find that amongst the radio quieter objects the radio emission is
strongly aligned with the optical minor axes. This requires there to
be negligible projection effects suggesting that the galaxies are
oblate, and possibly rotationally supported, with the jet emerging
along the rotation axis. Amongst the radio louder objects there is no
preferred radio-optical alignment, more akin to what we would expect
if the host galaxies were triaxial and not rotationally supported. We
suggest that this dichotomy in alignment behaviour is another
manifestation of the primarily optical dichotomy which exists between
giant ellipticals and normal (and dwarf) ellipticals (see for example
Faber et al., 1997; Kormendy et al. 2008). The former galaxies have
cores, are slowly rotating and triaxial, and generally have boxy
isophotes, while the latter are coreless, are rapid rotators and
oblate, and generally have disky isophotes. Assuming our association
of the two dichotomies is correct, a surprising consequence is that
the distinction in terms of mass between ``giant'' and ``normal''
elliptical galaxies for the subset with radio emission appears to be
very modest. In Fig.\ref{fig:mass} we show distributions of stellar
masses for the two subsets. The radio quieter objects are indeed less
massive but only by around a factor of two. Given that the observed
ratio of radio to stellar luminosity $(t - r)$ spans more than three
orders of magnitude, it is instructive to think in terms of the
efficiency with which the different types of galaxy produce radio
emission for a given stellar mass. From this viewpoint the isotropic,
rotationally supported galaxies are low efficiency radio emitters
while the anisotropic, slow rotators, have relatively high radio
efficiencies. Getting the gas from stellar mass loss to where it can
be accreted by the black hole is an obvious candidate that can affect
the efficiency. Alternatively, the black hole spin, which may
relate to the formation history of the galaxy, could affect the efficiency
with which accreted fuel is converted into radio emitting jets.

\section{Summary}

We find a highly statistically significant tendency for the axis of
the radio emission to align with the minor axis of the starlight in
elliptical galaxies. Surprisingly, this trend is confined to the less
radio-luminous objects. We suggest that this result may be a new
manifestation of the well known dichotomy in the optical properties of
elliptical galaxies; i.e. between those with cores and which are slow
rotators and those without cores and which are rotationally supported.

{}
\end{document}